\documentstyle[aps,epsf,twocolumn]{revtex}
\begin{document}
\draft
\title{Nonexponential Relaxation of Magnetization at the Resonant
Tunneling Point under a Fluctuating Random Noise}

\author{Seiji MIYASHITA and Keiji SAITO}

\address{
Department of Applied Physics, Graduate School of Engineering, \\
University of Tokyo, Bunkyo-ku, Tokyo 113-8656, Japan}

\date{\today}
\maketitle
\begin{abstract}
Nonexponential relaxation of magnetization at resonant tunneling points
of nanoscale molecular magnets is interpreted to be an effect
of fluctuating random field around the applied field.
We demonstrate such relaxation in Langevin equation analysis and 
clarify how the initial relaxation (square-root time) changes to
the exponential decay. The scaling properties of the relaxation
are also discussed.
\end{abstract}
\noindent
\vspace*{-0.1cm}

As to the relaxation of metastable magnetization of 
uniaxial nanoscale molecular magnets such as Mn$_{12}$
and Fe$_8$, the resonant tunneling phenomena have been
paid attention and various interesting properties of the
phenomena have been reported\cite{BB1,exp1,exp2,TART,exp5,expP,exp6,WWRS,WW}.
Properties of the phenomena have been also investigated in the theoretical 
side\cite{th1,th2,th3,th4,PS2,th6,th7,KN,miya95,miya96,DMSGG}.
We have studied such relaxation from a view point of 
nonadiabatic transition among their discrete energy structure due to a
finite number of degrees of freedom. 
There we studied the relation between the relaxation rate and the sweeping velocity.
Actually such relation has been utilized to estimate the energy gap in 
a recent experiment\cite{WWRS,WW}.
Because the tunneling gap is so small,
thermal environments have strong influence on the tunneling, which has been studied
extensively.

In the case of static field, there have been many experiments reporting that
nonexponential relaxation occurs at the resonant point.
That is, the initial relaxation can not be fitted by usual exponential 
function but is fitted well by a stretched exponential function with
the exponent near 0.5 or 
a square-root function\cite{exp6,Ohm98,sq-exp2,sq-exp3}.
If the field would be precisely fixed at the resonant point, we should see the 
coherent tunneling.
Thus the observations indicate that the decoherent effect plays an
important role.  
This nonexponential relaxation has been interpreted 
as a phenomenon due to the fact that the region of the 
field of the resonant tunneling is very narrow and small fluctuation 
can detune the resonance condition by Prokof'ev and Stamp\cite{PS2}.
They considered the distribution of the internal field which mainly 
consists of dipolar field from other molecules, and investigated its
time evolution. Combining the evolution of the distribution and the relation between
the steady state distribution of the dipolar field and 
the magnetization,
they found a square-root time initial relaxation and an exponential
relaxation in the late stage, which explains the overall dependence
of the experiments.

In the detailed observation on the distribution of internal field $P(\xi)$ in 
Fe$_8$\cite{WW}, 
the square-root dependence is found even if the initial magnetization
is zero, and 
change of the distribution appears only locally 
and it is not associated with the reforming total distribution of $P(\xi)$.
These features are not compatible with the mean-field type analysis of 
$P(\xi)$, and thus it seems necessary to consider more general 
mechanism of the square-root time initial relaxation.
Actually it has been pointed out \cite{PRL_Fe57} that 
the fast fluctuation of  hyperfine fluctuations are very
important in the local field dynamics.
In this Letter, we propose an alternate explanation of this phenomenon
as a general phenomenon at narrow resonant points with fluctuating field
using a Langevin equation approach, i.e., using the Ornstein-Uhlenbeck
process\cite{OU}. 

Here we consider a two-level system for the simplicity and its
Hamiltonian is given by 
\begin{equation}
{\cal H}=h_{\rm ext}(t)\sigma^z-{\Delta E\over2}\sigma_x,
\label{ham}
\end{equation}
where $\sigma_x$ and $\sigma_z$ are the Pauli matrices and 
$h_{\rm ext}(t)$ is a time-dependent external
field, and $\Delta E\over2$ denotes the transverse field which represents
the quantum fluctuation of the $z$ component of the spin.
We solve the Schr\"odinger equation of this system with time dependent
field $h_{\rm ext}(t)$ by applying the time evolution function
\begin{equation}
|t+\Delta t\rangle=\exp(-i{\cal H}(t))|t\rangle,
\label{tev}
\end{equation}
where the $h_{\rm ext}(t)$ is prepared by a Langevin equation.

Corresponding to the molecules of Mn$_{12}$ or Fe$_{8}$,
we consider that the resonant point is very narrow and is regarded as 
a point. The external field consists of the static part $h_0$ and a 
fluctuating part $h(t)$:
\begin{equation}
h_{\rm ext}(t)=h_0 + h(t).
\end{equation}
The fluctuating part is caused by independent changes of 
many magnetizations around the site.
Thus we consider that this part is an assembly of independent fluctuation
\begin{equation}
h (t) = h (0) +\int_0^t\sum_j\delta h_j(s)ds,
\label{hext00} 
\end{equation}
where $\delta h_j(s)$ is the change of a field from the $j$-th site ($h_j$)
at time $s$. Here we assume that $\delta h_j(s)$ is independent of time and
position, and we regard it as a white gaussian noise $\eta(s)$
\begin{equation}
\langle\eta(t)\rangle=0,\quad \langle\eta(t)\eta(s)\rangle=2D\delta(t-s).
\end{equation} 
Here the assumption of the gauss distribution is not essential but
we assume it for the convenience of analysis.
Then, the relation(\ref{hext00}) is written as
\begin{equation}
h (t)  = h (0) +\int_0^t\eta(s)ds=W(t),
\label{hext0} 
\end{equation} 
where $W$ is the Wiener process
\begin{equation}
\langle W(t)\rangle=0,\quad \langle W(t)W(s)\rangle=2D{\rm Min}(t,s).
\end{equation} 
The distribution of $W(t)$ is given by
\begin{equation}
P(W)={1\over\sqrt{4\pi Dt}}\exp\left(-{W^2\over 4Dt}\right).
\label{WP}
\end{equation}
We show an example of the Wiener process in Fig. 1 (a dashed line).

The nonadiabatic transition only occurs when the field crosses the 
resonant point. Hereafter we set the resonant point at $h=0$ and 
also the static field at $h_0=0$.
The probability $p$ of the adiabatic transition of the state,
i.e., from the ground state to the ground state 
(or from the excited state to the excited state), is given by the 
Landau-Zener-St\"uckelberg (LZS) formula\cite{Landau,Zener,St}
\begin{equation}
p=1-\exp\left(-{(\Delta E)^2\over 2|M-M'|v}\right),
\label{p}
\end{equation}
where $\Delta E$ is the energy gap at the crossing point (tunnel gap),
$M$ and $M'$ are the magnetizations of crossing states, and $v$ is the   
speed of the field at the resonant point.
The transition probability from the initial distribution $(p_1, p_2)$
to the scattered distribution $(p'_1, p'_2)$ 
depends on the velocity $v$ of the field at the time of crossing  and 
also on the
phase factor due to the free motions outside of the crossing region\cite{teranishi-nakamura}. 
Here we consider many sample of $\{h_{\rm ext}(t)\}$ and the ensemble
average over the distribution of $v$ and the phase factor.

The change of magnetization $(M\rightarrow M')$ 
at this crossing is given by
$M=M_0(p_1-p_2) \rightarrow M'=M'_0(p'_1-p'_2)$, 
where $M_0$ and $M'_0$ are the magnetization of the each ground state.
In Fig. 1, we show a dynamics of magnetization $M(t)$ 
for the shown process of $h_{\rm ext}$.

From the dependence of (\ref{p}), the average probability of
the adiabatic transition, which
changes the magnetization at a crossing, is estimated as
\begin{equation}
p_{\rm av}\simeq \langle {(\Delta E)^2\over 2|M-M'|v} \rangle
=\alpha_0{(\Delta E)^2\over \sqrt{D}},
\end{equation} 
where we assumed that the value of $p$ is small and that the
average velocity of the field is proportional to the strength of the
random field $\sqrt{D}$.
The change of magnetization at a crossing is given by
$M'=(1-2p_{\rm av})M$.
\begin{figure}
\noindent
\centering
\epsfxsize=9.0cm \epsfysize=6.0cm \epsfbox{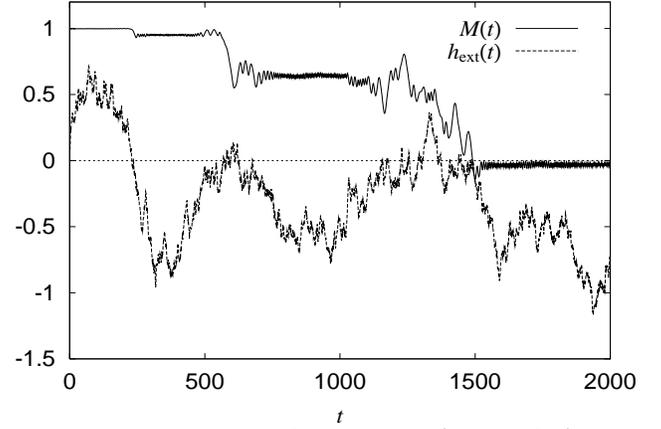} \\
\caption{A sample of the Wiener process (dashed line) and the 
change of the magnetization for this process of $h_{\rm ext}$.}
\end{figure}

\noindent
If we assume that the sequence of the crossings give independent
contributions, 
the change of magnetization for arbitrary initial value $M(0)$ is given by
\begin{equation}
M(t)=M(0)(1-2p_{\rm av})^{N_{\rm ac}(t)},
\end{equation} 
where $N(t)_{\rm ac}$ is the accumulated number of the crossings by the time of $t$.

The number of crossings per unit time, $N(t)$, is estimated from 
the probability for $|W|\le\sqrt{D}$, where $\sqrt{D}$ is the jump range
of the field. Thus $N(t)$ at time $t$ is given by   
\begin{equation}
N(t)\propto{1\over\sqrt{4Dt}}\times \sqrt{D} \propto{1\over\sqrt{t}}.
\end{equation}
Thus the  accumulated number of crossings by the time $t$ is given by
\begin{equation}
N_{\rm ac}(t)=\int_0^tN (t)dt=c\sqrt{{t}},
\end{equation}
where $c$ is a constant.
Thus the total change of the magnetization is
$$M=M_0(1-2\alpha_0{(\Delta E)^2\over \sqrt{D}})^{c\sqrt{t}}$$
\begin{equation}
\simeq \exp\left(-\alpha{(\Delta E)^2\over \sqrt{D}}\sqrt{t}\right),
\label{M-strech}
\end{equation}
where $\alpha=2\alpha_0c$. This constant is uniquely determined by the 
nature of random process which will be determined later.
Thus we conclude that the magnetization shows a stretched exponential decay
with the exponent $1/2$, which shows the square-root time initial relaxation.
In Fig. 2, we show the averaged magnetization $\langle M(t)\rangle$ 
over the 10,000 samples of $h_{\rm ext}$. In the inset $M(t)$ is plotted 
in the coordinate $(\sqrt{t}, \log M(t))$, where we confirm the 
dependence of (\ref{M-strech}).
This dependence gives the initial square-root time dependence
\begin{equation}
M=M_0(1-\alpha{(\Delta E)^2\over \sqrt{D}}\sqrt{t}).
\label{m-initial}
\end{equation} 
\begin{figure}
\noindent
\centering
\epsfxsize=9.0cm \epsfysize=6.0cm \epsfbox{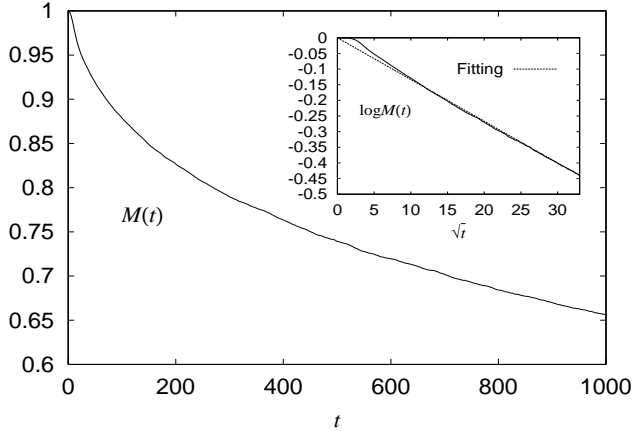}
\caption{The averaged magnetization $\langle M(t)\rangle$ 
over the 10,000 samples of $h_{\rm ext}$. The inset shows a plot 
$( \log M(t), t^{1/2} )$, where the dotted line is a straight guide line
for eye.}
\end{figure}

In the real situation the field $h(t)$ may not deviate without limit and
there are some restoring mechanisms. 
Next we study the mechanism of transition from the initial relaxation (\ref{M-strech})
to the exponential relaxation.
Taking into account the restoring effect, we adopt the Ornstein-Uhlenbeck process
for the evolution of $h(t)$:
\begin{equation}
{dh\over dt}=-\gamma h(t)+\eta(t).
\label{OUP}
\end{equation}
The process (\ref{hext0}) corresponds to the case $\gamma=0$.
The distribution of $h(t)$ for this process is given by
\begin{equation} 
P(h)={1\over\sqrt{2\pi\sigma^2}}\exp\left(-{(h-\langle h\rangle)^2\over 
2\sigma^2}\right),
\end{equation}
where
\begin{equation}
\sigma^2={D\over\gamma}\left[1-\exp(-2\gamma t)\right].
\end{equation}
For $t\ll \gamma^{-1}$, this reproduces (\ref{WP}). 
On the other hand it has a stationary distribution $\sigma^2=D/\gamma$  
at large time $t\gg \gamma^{-1}$.  
At this late stage there is a constant probability that
the $h(t)$ stays around 0, which causes a constant rate relaxation,
i.e., the exponential relaxation.
Therefore we expect 
\begin{equation}
M=M_1\exp\left({-\beta(\Delta E)^2\sqrt{\gamma\over D} t}\right),
\label{M1}
\end{equation}
for a long time $t\gg (2\gamma)^{-1}$, 
while we have the relation (\ref{M-strech})
for a short time $t\ll (2\gamma)^{-1}$.
Here $\beta$ is a constant independent of $\Gamma$, $\gamma$,
and $D$. $M_0$ and $M_1$ are constants corresponding to 
a kind of initial magnetization of each process, which does not 
coincide with the initial magnetization. 

In Fig. 3(a) we show a time dependence of the magnetization
$\langle M(t)\rangle$ averaged over 10,000 samples.
In Figs. 3(b) and (c), we plot $M(t)$ in $(\sqrt{t},\log M(t))$ and
in $(t,\log M(t))$, respectively. There
we find the crossover from the square-root time relaxation to the 
exponential relaxation around $t\sim (2\gamma)^{-1}$. 
\begin{figure}
\noindent
\epsfxsize=9.0cm \epsfysize=10.0cm \epsfbox{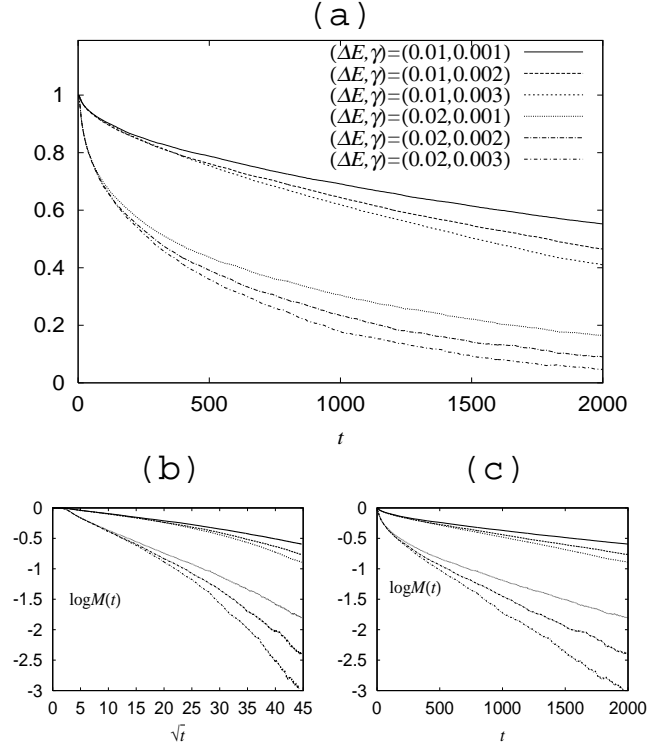} \\
\caption{Parameter dependence of $\langle M(t)\rangle$
for $h_{\rm ext}$ given by (\ref{OUP}) for $D=0.001$:
(a) the linear plot $(t,\langle M(t)\rangle)$,
(b) plot $(\sqrt{t},\log\langle M(t)\rangle)$ for the initial relaxation (\ref{M-strech}),
and 
(c)  plot $(t,\log\langle M(t)\rangle)$ for the late relaxation (\ref{M1}).
}
\end{figure}

In a very early time $t<5$ we find a dead time where the magnetization
does not change. This is considered to be due to the fact that
the noise starting at 0 stays near the resonant region where the system
essentially evolves coherently, i.e. $M(t)\simeq\cos(\Delta Et)$.
This dead time becomes relatively short when $\Delta E$ and $\gamma$
becomes small.

Studying $\langle M(t)\rangle$  for various parameters $(\Delta E, \gamma, D)$
we confirm that scaling relations of $M(t)$ on the parameters 
indicated by (\ref{M-strech}) and (\ref{M1}).
In Figs. 4(a) and (b), we plot the data  
($\Delta E=0.01$, $\gamma=0.001,0.002$, and 0.003, and $D=0.001, 0.002$ and 0.003)
in the coordinates:
$(\sqrt{t},\log M(t)/((\Delta E)^2/\sqrt{D}))$ and
$(((\Delta E)^2 \sqrt{\gamma/D})t,\log M(t))$, respectively.
We could plot the data in $( ((\Delta E)^2/\sqrt{D})\sqrt{t},\log M(t))$ 
instead of Fig. 4(a).
However the scaling time region 
$[{\rm the \ dead \ time}\sim O(1) < t < (2\gamma)^{-1}]$ 
of each parameter set
shown in different region of $((\Delta E)^2/\sqrt{D})\sqrt{t}$ and it looks mess. 
Thus we plot data in the way of Fig. 4(a).  
There the lines shows some distribution due to the distribution of $M_0$ and $M_1$,
and also the effect of the dead time. However, we find that 
slopes of them are almost 
the same to each other and we estimate the constants as
\begin{equation}
\alpha_0\simeq 3.5\pm 0.1,\quad {\rm and }\quad \beta_0\simeq 2.3\pm0.1.
\end{equation}
When we change $\Delta E$, the value of  $M_0$ and $M_1$ and the dead time change,
but we find that the values of $\alpha$ and $\beta$ are consistent 
although the scattering of date is larger.
\begin{figure}
\noindent
\centering
\epsfxsize=8.0cm \epsfysize=10.0cm \epsfbox{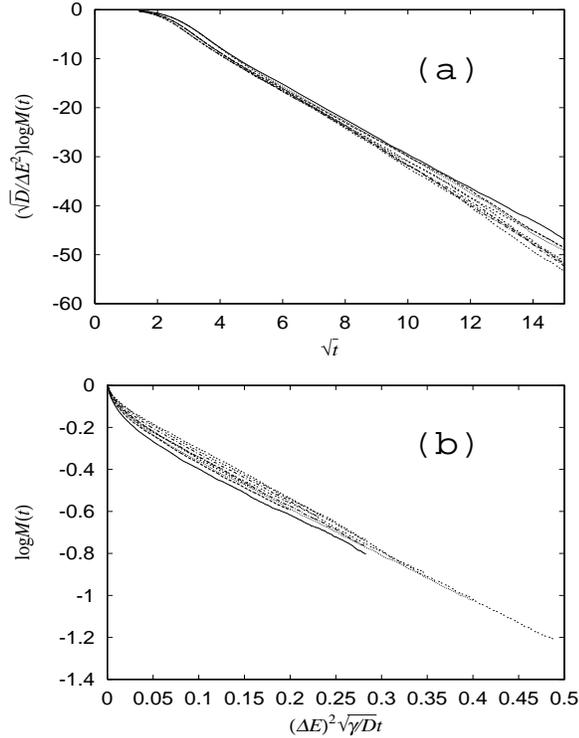} \\
\caption{Scaling plots (a)$(\sqrt{t},\log M(t)/(\Delta E^2/\sqrt{D}))$,  
and (b)$( ( (\Delta E)^2 \sqrt{ \gamma / D} ) t,\log M(t))$}
\end{figure}

Finally we consider the case where the property of the noise depends 
on $M(t)$. Generally the change of the average magnetization causes the
change of distribution of the field $h_{\rm ext}$ as discussed 
by Prokof'ev and Stamp\cite{PS2}. In our analysis this change should be taken into account
as a slow change of $h_0$, which may lead the same mechanism discussed in \cite{PS2}, which
will be discussed in the future. Effect of change of the mean field to the 
LZS transition has been also discussed as a feedback effect which 
causes a large change of the transition probability\cite{feed-back}. 
This effect also causes important modification when the field is
swept where the effective sweeping rate is modified.
In the present case of fixed field, we assume that
the time scale of the fluctuation field is much smaller
than the change of the average magnetization. 

We would like to thank Bernard Barbara and Wolfgang Wernsdorfer for their valuable
discussions. The present work is partially supported by the Grant-in-Aid
from the Ministry of Education. 

\end{document}